# Aerial Robot Model based design and verification of the single and multi-agent inspection application development

Seiko P. Yamaguchi[1a,2], Masaru Sakuma[1a], Takaki Ueno[1a], Filip Karolonek[2]
Tadeusz Uhl[2], Ankit A. Ravankar[1b], Takanori Emaru[1b], Yukinori Kobayashi[1b*]

[1] Laboratory of Robotics & Dynamics, Div. of Human Mechanical Systems and Design
[a] *Graduate School* / [b] *Faculty* of Engineering, Hokkaido University, Sapporo, Japan
[2] Department of Robotics and Mechatronics, Faculty of Mechanical Engineering and Robotics,
AGH University of Science and Technology, Cracow, Poland
**spymgc@eis.hokudai.ac.jp, *kobay@hokudai.ac.jp**

**Abstract**

In recent decade, potential application of Unmanned Aerial Vehicles (UAV) has enabled replacement of various operations in hard-to-access areas, such as: inspection, surveillance or search and rescue applications in challenging and complex environments. Furthermore, aerial robotics application with multi-agent systems are anticipated to further extend its potential. However, one of the major difficulties in aerial robotics applications is the testing of the elaborated system within safety concerns, especially when multiple agents are simultaneously applied. Thus, virtual prototyping and simulation-based development can serve in development, assessment and improvement of the aerial robot applications. In this research, two examples of the specific applications are highlighted: harbor structure and facilities inspection with UAV, and development of autonomous positioning of multi-UAVs communication relaying system. In this research, virtual prototype was designed and further simulated in multi-body simulation (MBS) feigning the sensing and actuating equipment behaviors. Simultaneous simulation of the control and application system running with software in the loop (SITL) method is utilized to assess the designed hardware behavior with modular application nodes running in Robot Operating System. Furthermore, prepared simulation environment is assessed with multi-agent system, proposed in previous research with autonomous position control of communication relaying system. Application of the virtual prototype's simulation environment enables further examination of the proposed system within comparison degree with postfield tests. In future work, the proposed simulation environment can be further applied for motion and vibration control of the UAV applications. The research aims to contribute through case assessment of the design process to safer, time and cost-efficient development and application design in the field of aerial robotics.

## 1. Introduction

Evolution and development of UAVs have enabled their utilization in variety of robotic operations [1] [2]. Although, enormous potential of aerial robots exists, high concern to safety and regulatory restrictions is a major bottleneck during system design and development phase. Hence, the article presents the design procedure with mechatronic approach, verified and confirmed to be effective and time-efficient for innovative product and system design [3]. Virtual prototyping and verification with simulation tools decreases the risk of damage while failures. In this research, two cases of aerial robots anticipated applications are concerned, presenting the system and software design process with mechatronic model-based approach of: hard-to-access area structure inspections, and multi-UAV communication relaying [4].

## 2. Problem Formulation

As one of the challenges considered, operation range of aerial platforms are limited due to its wireless communication. Increasing operation range of UAV missions can be achieved by multi-agent communication relaying, anticipated to serve in better endurance of the system applications [5] (Fig.1). However, multi-UAV system design where several agents are simultaneously applied in air faces a high risk of damage during failures. Hence, virtual-prototype and simulation-based development should decrease design time and improve its reliability before the hardware laboratory model is tested.

Application requirements considering monitoring of port facilities including pier structure and breakwater monitoring had been revised (Fig.2). Challenges of UAV utilization in this area arise due to several reasons; Wide area coverage and long-range flight required for a single mission, hard-to-access environment as well as possibility of GPS-denied missions. The large aspect of the wide requirements needs reflection in application development.

### 3. Virtual Prototyping and Simulation based Design

Aerial robotic system includes UAV platform, sensing and actuating components, flight controller and computational responsibility. Fig.3(i) presents the set-up of the aerial robot. Virtual prototype of the UAV's rigid body model was developed with CAD and converted to simulation supported format to be applied for multi-body simulations. Thus, integrating the simulation model with flight simulator and software operating system, one can imitate the behavior of the aerial robot within SITL simulation, as in Fig.3(ii). In combination, HIL setup is enabled prior to field tests.

### 4. Implementation and Verification

Model based designing process was applied to following application design and development with proposed methods to highlight its strength.

#### 4.1 Multi-UAV Communication Relaying

System structure as well as autonomous positioning system for formation control [4] was developed and verified. Fig.4(i) presents dynamic MBS simulation interfacing with system developed, Fig.4(ii) formation control monitoring with SITL simulation. It is remarkable that during the software design of the fail-safe collision avoidance between UAVs, proposed system enabled to verify algorithm developed with no hardware damaged due to SITL tests proceeded within software development stage.

#### 4.2 Harbor Inspection Scenarios

In order to test capabilities in challenging environments, application development for data acquisition, localization and navigation methods were considered. Modular experimental possibilities of the simulation environment accelerated the choice of components and verification for application building. For example, 3D point-cloud data accusation and processing presented in Fig.5, or SLAM [6] and navigation techniques [7] for GPS-denied environment [8] [9].

### 5. Conclusion and Future Work

The applied method has presented its effectiveness in acceleration of design process and prototyping, as well as increased reliability in safety. Although, further verification of the simulation accuracy is required, communication relaying had been developed with this method. Thus, practical challenges highlighted during actual hardware tests are brought over to improvements with model-based simulations, including necessary improvement in simulations. Harbor inspection application is still in the development phase. Advantages of simulation-based development was highlighted due to the access difficulties to aimed structures. The research plans to further verify the solution developed with testbed implementations.

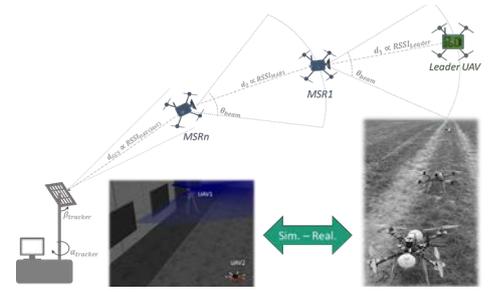

*Figure 1: Operation principle of the Multi-UAV communication relaying and its system design through model based simulations, followed by hardware tests.*

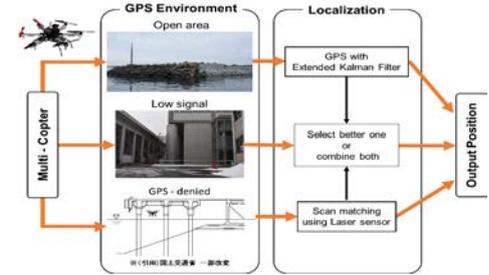

*Figure 2: Multi-copter requirements on port facilities inspection. System design process should concern wide requirements characteristics within its application.*

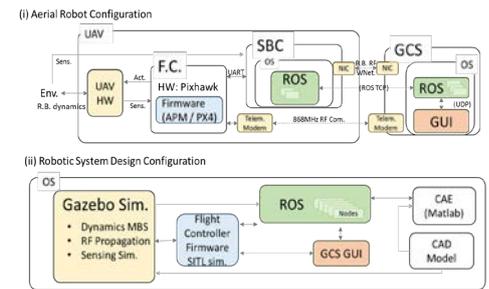

*Figure 3: System diagram comparison of aerial robot components and model-based simulation environment*

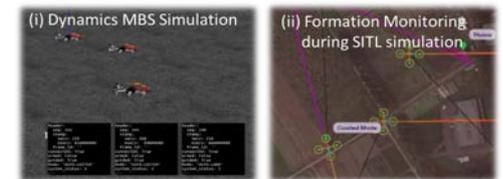

*Figure 4: Multi-UAV Communication relaying system design and verification with simulation environment,*

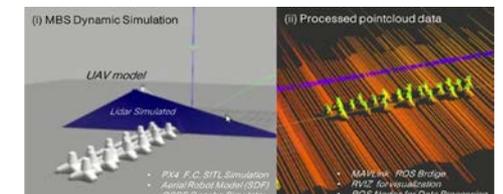

*Figure 5: Simulation of the UAV-based breakwater 3D point-cloud acquisition with 2D lidar and flight data.*